# On relative contribution of electrostatic and aerodynamic effects to dynamics of a levitating droplet cluster

Alexander A. Fedorets [a], Leonid A. Dombrovsky [a,b,1],

Edward Bormashenko [c], and Michael Nosonovsky [a,d]

[a] University of Tyumen, 6 Volodarskogo St, Tyumen, 625003, Russia

[b] Joint Institute for High Temperatures, 17A Krasnokazarmennaya St, Moscow, 111116, Russia

[c] Department of Chemical Engineering, Biotechnology and Materials, Engineering Science Faculty, Ariel University, Ariel, 40700, Israel

[d] Department of Mechanical Engineering, University of Wisconsin–Milwaukee, 3200 North Cramer St, Milwaukee, WI 53211, USA

**Abstract**

New experimental results and their physical analysis are presented to clarify the behavior of a relatively stable self-arranged droplet cluster levitating over the locally heated water surface. An external electric field of both opposite directions leads to a significant increase in the rate of a condensational growth of droplets in the cluster. The experimental data are used to estimate a small electrical charge of single droplets and the attraction force of polarized droplets to the water layer. It is confirmed that the interaction between the droplets is governed by aerodynamic forces.

**Nomenclature**

| | | | |
|---|---|---|---|
| $E$ | electric field strength | $\dot{S}$ | rate of increase of droplet surface area |
| $e$ | elementary charge | $t$ | current time |
| $H$ | height of cluster/droplet levitation (Fig. 1) | $U^{\pm}$ | two configurations of electric field and voltage at the upper electrode |
| $h$ | distance between droplet and water layer (Fig. 1) | | |
| $L$ | distance between centers of neighboring droplets | | |
| $l_U$ | distance between electrodes | $u$ | velocity |
| $q$ | electric charge | $W$ | power |
| $R$ | radius of droplet | $z$ | axial coordinate |
| $r$ | radial coordinate | | |
| $S$ | area of droplet surface | | |

---
[1] Corresponding author. Tel. +7 910 408 0186
Email address: ldombr@yandex.ru

| *Greek symbols* | | *Subscripts and superscripts* | |
|---|---|---|---|
| $\gamma$ | coefficient in Eq. (7) | ad | aerodynamic |
| $\varepsilon$ | dielectric constant | b | boundary |
| $\eta$ | dynamic viscosity | c | coalescence, critical |
| $\xi$ | coefficient in Eq. (8) | el | electrostatic |
| $\rho$ | density | dp | dipole |
| $\varphi$ | electric potential | dr | droplet |
| | | L | laser |
| | | max | maximum |
| | | z | axial |

## 1. Introduction

This particular study is a continuation of an experimental research and physical modeling self-assembled levitating clusters of droplets generated over the locally heated water surface. The previous stages of a long-time research initiated by the discovery of this fascinated phenomenon [1] have been recently described in keynote presentation [2]. It is clear at the moment that aerodynamic forces determine both the formation and evolution of the droplet cluster [2–4]. However, electric charging the droplets during their evaporation [5, 6] needs independent estimates to clarify a contribution of Coulomb forces to the quasi-steady cluster parameters. In addition, the effect of an external electric field on a droplet cluster behavior is expected to be interesting for the study of a droplet cluster stability. The above reasons motivated the present study. The objective of the paper is two-fold: (1) To estimate the own electrical charge of single water droplets in a droplet cluster and the resulting forces between the droplets and (2) To use new experimental data for the effect of an external electric field to explain the main parameters of a stable droplet cluster. It should be noted that the life of a cluster is usually not long (just a few tens of seconds) because of condensational growing of droplets and the resulting coalescence of large droplets with the layer of water. So, the cluster stabilization is one of the key problems [7–10].

## 2. Experimental procedure

The experimental equipment used to produce droplet clusters was described in detail in [10, 11]. The cluster of water droplets was formed over a thin layer of distilled water heated locally from the solid substrate irradiated from below by a laser beam of diameter about 1 mm. The continuous wave laser KLM-H808-600-5 with the wavelength of $0.808\,\mu m$ at the working power of $W_L = 280\,mW$ was used in all the experiments. The thickness of water layer was equal to $400 \pm 2$ μm in all experiments. It was controlled using the confocal chromatic sensor IFC2451 made by the company Micro-Epsilon (USA). Video images of the cluster were taken using stereomicroscope Zeiss AXIO Zoom. V16 and high-speed video-camera PCO.EDGE 5.5C (Germany) with spatial resolution of 0.6 μm.

Both the design and position of electrodes are schematically shown in Fig. 1. The lower electrode is a thin metal cylinder (outer diameter – 8.9 mm, the diameter of central orifice 4.5 mm, and thickness

– 0.7 mm) placed directly under the substrate. The upper electrode with outer diameter 78 mm and central hole with diameter 6.5 mm was made of fiberglass laminate covered by 18 μm thick copper layer. The distance between the electrodes was equal to $l_U = 8$ mm. The external electric field was generated by high-voltage source HVLAB3000 (ET Enterprises, UK), which is designed to vary the applied voltage in the range from 0.2 to 3.0 kV. For convenience, the variant with a positive electric potential of the upper electrode relative to the grounded lower electrode will be denoted by $U^+$, and the opposite configuration – $U^-$.

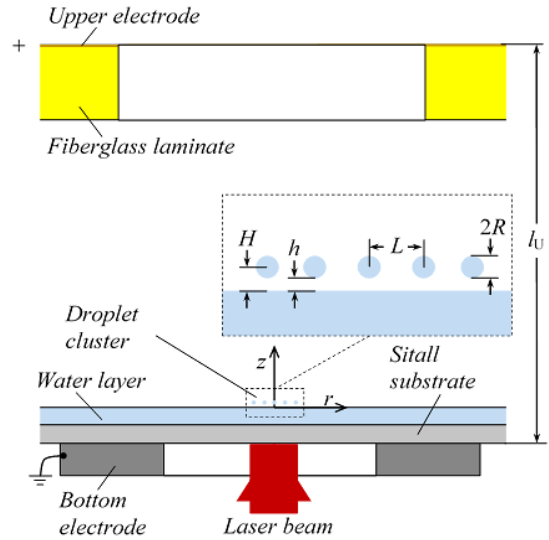

Fig. 1. Schematics of a side view of a laboratory set-up.

## 3. External electric field and polarization of water droplets

Before proceeding to experimental results, it seems correct to calculate the electric field in the location of droplet cluster. This can be done by a numerical solution of the axisymmetric boundary-value problem for the Laplace equation for electric potential $\varphi(r, z)$:

$$\nabla(\varepsilon \nabla \varphi) = 0 \qquad (1)$$

where $\varepsilon(r, z)$ is the permittivity of substances. Note that distilled water used in the experiments is really a dielectric (in contrast to water with some impurities). The values of $\varepsilon = 81$ for water and $\varepsilon = 6$ for the substrate substance were used in the calculations. The obvious boundary conditions for electric potential, $\varphi = \varphi_b$ or $\partial \varphi / \partial n = 0$, are used at different parts of the region boundary. The finite element method (FEM) was employed in the calculations [12, 13]. A non-uniform FEM mesh with 3200 triangular elements is definitely sufficient for accurate calculations. The numerical results obtained for $U^+ = 1$ kV are presented in Fig. 2. The electric field in the location of cluster is uniform, and the absolute value of field strength, $E = \partial \varphi_z / \partial z$ where $\varphi_z = \varphi(0, z)$, is equal to $220$ kV/m. The linearity of the problem enables us not to repeat the calculations at various values of $U^+$ or $U^-$.

Water droplets of a cluster are polarized in the electric field. The absolute value of dipole moment of a single spherical droplet with a radius $R$ in a uniform external field can be calculated as follows [14]:

$$P = 4\pi\varepsilon_0 \frac{\varepsilon-1}{\varepsilon+2} R^3 E \qquad (2)$$

where $\varepsilon_0 = 8.85 \cdot 10^{-12}$ F/m is the vacuum permittivity. The surface density of electric charge of a polarized droplet is proportional to the cosine of the angle, measured from the vertical axis. Of course, the charge of the upper and lower surfaces of the droplet is the same in absolute value and opposite in sign.

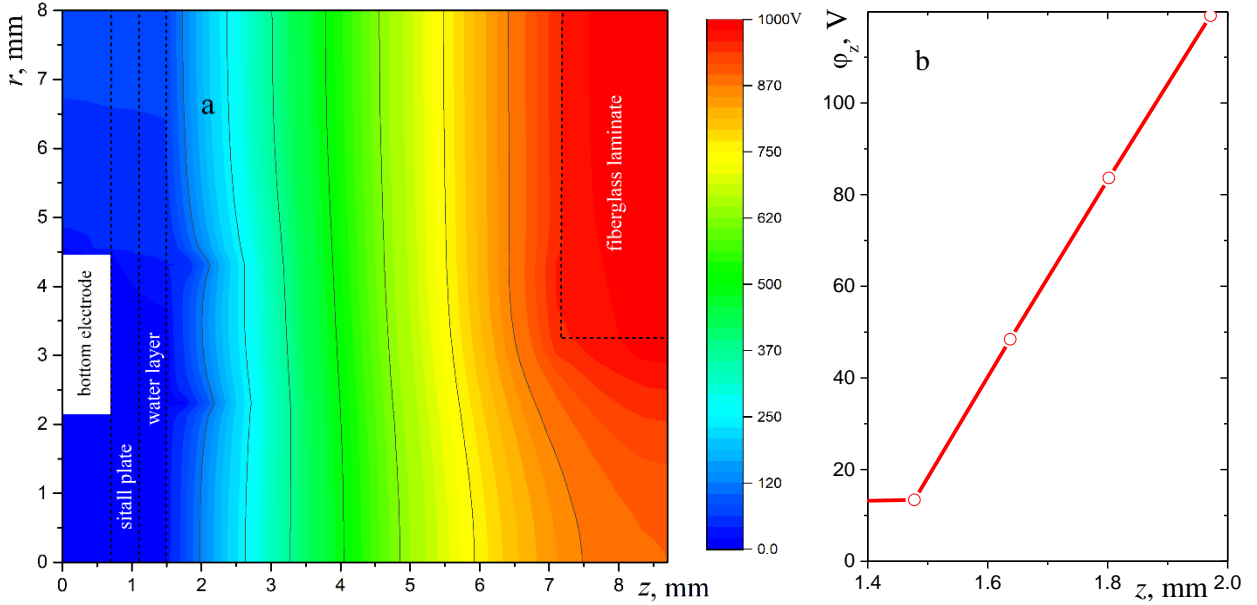

Fig. 2. The field of electric potential: a – in the computational region,

b – in the cluster location (along the axis $r = 0$).

## 4. Experimental results on behavior of droplet cluster

The new laboratory experiments showed a significant effect of external electric field on coalescence of droplet cluster with the water later. At ordinary conditions, without any external electric field, the coalescence of droplet cluster has been studies in detail [15–17]. The distance $h$ between droplets and water layer decreases with the condensational growth of water droplets. As a result, one of the largest droplets touches the surface of water layer and generates capillary waves which lead to avalanche-like coalescence of other droplets. This process changes radically with an external electric field. Coalescence begins with a significantly smaller droplet size. This process is illustrated in Fig. 3: the droplet and its reflection circled in a dashed ellipse in Fig. 3b disappear in Fig. 3c. The coalescence of a single droplet generates high-frequency capillary waves of small amplitude without any effect on the neighboring droplets. The coalescence appears to be similar for both $U^+$ and $U^-$ configurations, but the critical droplet radius, $R_c$, is greater for the $U^+$ variant. The time dependences of

$h(t-t_c)$ are plotted in Fig. 4. Two stages of the process are quite clear: the relatively slow uniform decrease in the height of the water droplet levitation is replaced by the sharp decrease in approximately 0.05 s before the coalescence.

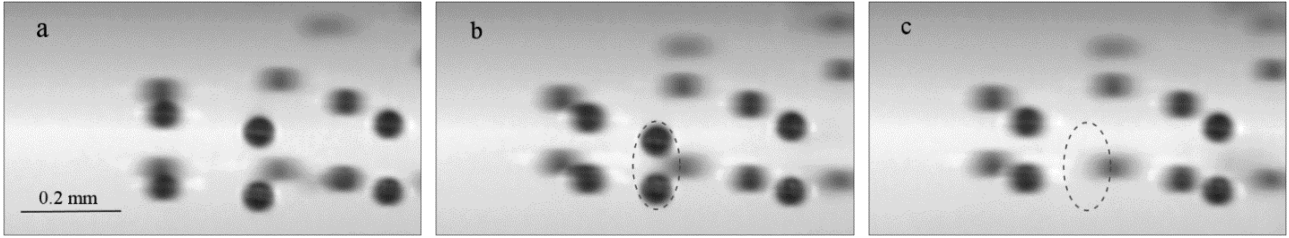

Fig. 3. Dynamics of coalescence of a droplet with water layer at $U^+ = 1\text{kV}$:
a – $t = t_c - 0.5\text{s}$, b – $t = t_c - 0.01\text{s}$, c – $t = t_c$.

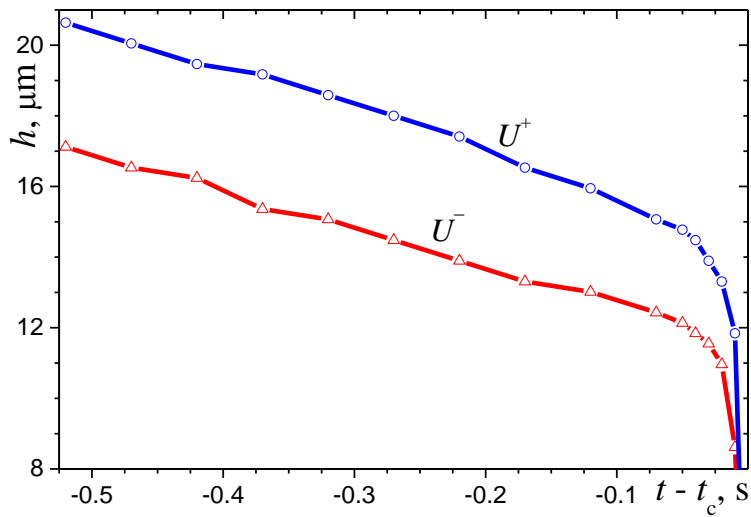

Fig. 4. Time variation of the distance between the droplet and water layer at $U^\pm = 1\text{kV}$.

In contrast to the coalescence process, the structure of droplet cluster is weakly sensitive to the external electric field. A comparison of the cluster patterns at $U^\pm \leq 0.7$ (to avoid the coalescence) shows only minor changes of distances between the droplets. The detailed measurements for the central part of cluster at $U^\pm = 0.6\,\text{kV}$ showed only about 1 % difference in the values of $L$ as compared with the case of $U = 0$. This unexpected result will be discussed below.

It is known that the rate of droplet growth due to condensation at ordinary conditions (without an external electric field) is well described by the d-squared law [18] or its modification called the elliptic law [19]. Therefore, according to previous papers by the authors, the value of the surface area rate of growth, $\dot{S}$, is used to analyze the effect of electric field. As earlier, the measurements were performed for the central part of the droplet cluster [8, 10, 11, 20]. The new experimental data are presented in Fig. 5 in the form of ratio of $\dot{S}/\dot{S}_0$ where $\dot{S}_0$ was obtained in the experiment without electric field. The general trend of increasing the rate of growth of water droplets with the electric field intensity is quite obvious. It is interesting that the direction of electric strength vector appeared to be an important factor, whereas the qualitative effect is the same in both configurations of electric field. The symbols $U^+$ and $U^-$ in Fig. 5 correspond to different signs of the electric potential of the upper electrode, whereas the potential of the bottom electrode is equal to zero.

One can expect that relatively large growth rate of water droplets in experiments with an external electric field is explained by the increase in evaporation rate of water layer. It was shown in early paper [21] that distilled water evaporated much faster in the presence of a high-voltage electric field in the direction normal to the water surface. Some details about the "Asakawa effect" and its applications can be found in [22–25]. Unfortunately, it is difficult to measure local increase in water evaporation in a small hot region just below the cluster. At the same time, the laboratory measurements of total evaporation from the whole water layer didn't indicate any considerable effect of the electric field. Perhaps, the increase in local evaporation rate of a hot water takes place but it is partially compensated by a continuous coalescence of relatively small droplets with the water layer.

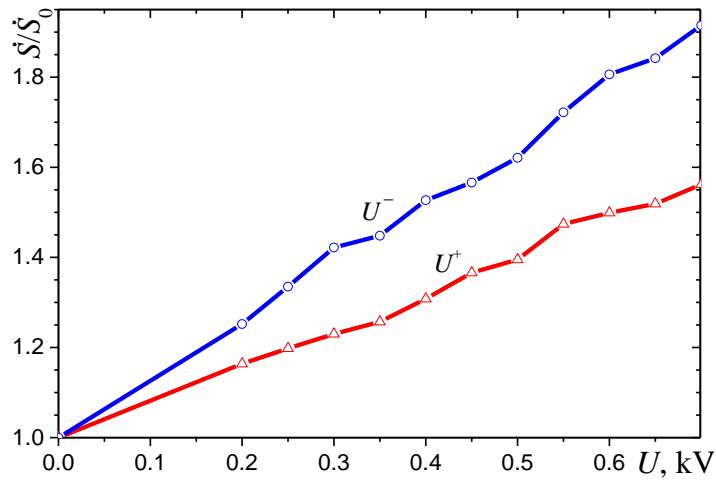

Fig. 5. Effect of electric field on growth rate of water droplets

## 5. Analysis of experimental data

The laboratory experiments showed that the external electric field decreases the critical size of water droplets at the coalescence beginning. The monotonic dependences of $R_c$ on the local electric field (see Fig. 6) indicate that the difference between the data for $U^+$ and $U^-$ is relatively small as compared with the "$U^\pm$" effect. This can be treated as an insignificant role of the negative electric charge of droplets as compared with the interaction of polarized dielectric droplets with water layer. Note that there is no contradiction between the relatively fast growth of droplets in the case of $U^-$ configuration (Fig. 5) and the lower values of $R_c$ (Fig. 6). The latter is explained by an additional force attracting the negatively charged droplets to the positively charged surface of water layer.

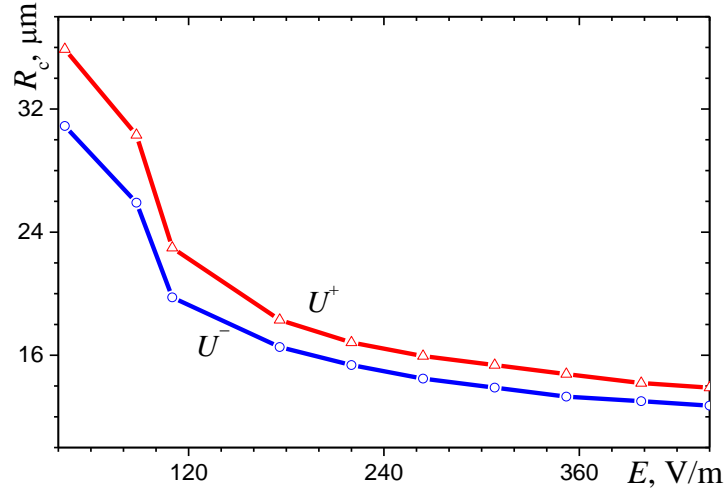

Fig. 6. Effect of electric field on critical radius of water droplets.

It is interesting to consider the dependences of weight of steady levitating small droplets on the droplet height of levitation at rather strong external electric field of opposite directions. Such dependences together with that for the case without the electric field are presented in Fig. 7. Note that the droplet weight range in this figure corresponds to the narrow radius range of $14.8 < R < 18\,\mu m$.

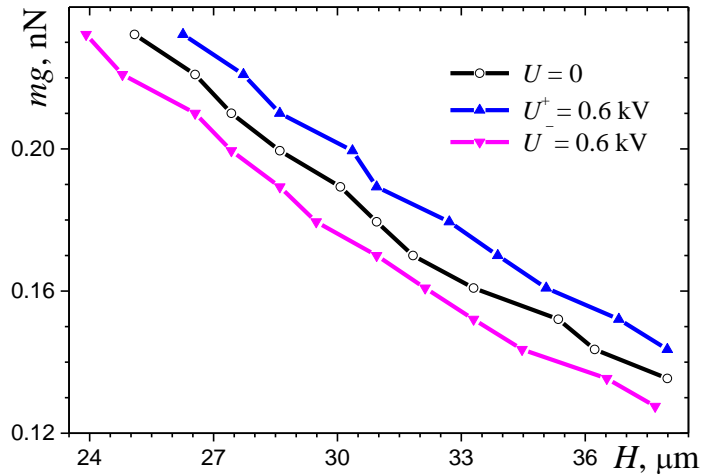

Fig. 7. The weight of a steady levitating droplet as a function of the levitation height.

Three curves in Fig. 7 are almost parallel to each other and the upper and lower curves are practically symmetric with respect to the central one with the ordinate distances between the neighboring lines $\Delta(mg) \approx 0.01\,\text{nN}$. It is clear that the electric charge of all the droplets is practically the same for water droplets under consideration and can be calculated as follows:

$$q_{dr} = \Delta(mg)/E(U^{\pm}) \approx 473\,e \qquad (3)$$

where $e$ is the elementary unit charge. This value of droplet charge is in good agreement with an estimate of $q_{dr} \sim 10^2 - 10^3\,e$ obtained in papers [5, 6] using quite different methods. It should be recalled that charged water droplets are the ordinary objects in the lower atmosphere [26–29]. So, the laboratory studies of the effect of electrical charging on behavior of small droplets and their clusters are also interesting for geophysical applications.

An estimate of the attraction force between a polarized water droplet and the water layer can be obtained assuming that the opposite charges of the dipole are located at the upper and lower points of the droplet surface. Obviously, the effect of upper charge can be ignored at small values of $h$ comparable to the droplet radius. To simplify the estimate, one can also assume that the electric charge induced upon the water layer surface is equal in absolute value to the dipole charges and located just below the droplet. These assumptions enable us to use the following equations for the absolute value of dipole electric charges and the attracting force between the dipole and water layer:

$$q_{dp} = \frac{P}{2R} = 2\pi\varepsilon_0 \frac{\varepsilon-1}{\varepsilon+2} R^2 E \qquad F_{dp}^{max} = \frac{1}{4\pi\varepsilon_0} \frac{q_{dp}^2}{h^2} = \pi\varepsilon_0 \left(\frac{\varepsilon-1}{\varepsilon+2}\right)^2 E^2 \frac{R^4}{h^2} \qquad (4)$$

The calculated dependences of $F_{dp}^{max}(h)$ presented in Fig. 8 indicate a strong increase of the attracting force with decreasing the distance between typical polarized droplets and a layer of water. It is important that $F_{dp}^{max}$ is comparable to the gravitational force at realistic values of $h$ (see Fig. 7). In the case of a stable cluster, the greatest aerodynamic drag force is equal to the sum of the gravitational force and the discussed electric attracting force.

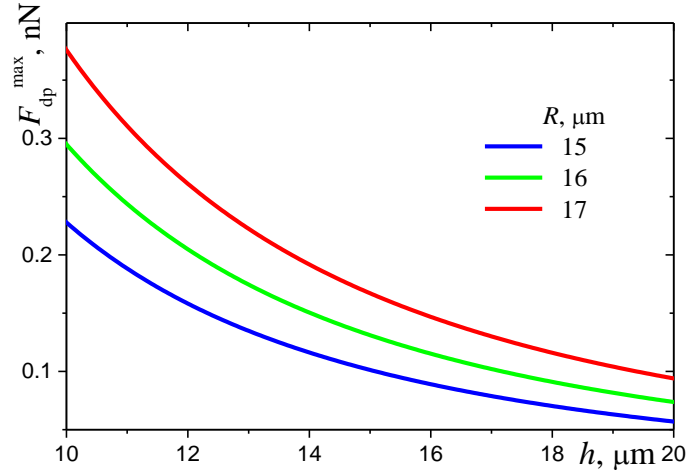

Fig. 8. The electric force attracting polarized water droplets to substrate water layer.

Let us proceed to the most interesting problem of a regular and relatively stable structure of a self-arranged droplet cluster observed in the experiments. It was noted above that the values of $L$ at $U^{\pm} = 0.6\,\text{kV}$ are practically the same as that without the electric field. One can say that additional forces between the neighboring droplets due to their polarization are very small and cannot affect the cluster structure. However, this argument is insufficient to explain the small effect of electric field. It is interesting to recall that the distances between droplets in the ordinary cluster (without an external electric field) are very sensitive to the steam generation rate controlled by the laser heating power. The latter statement is clearly illustrated in Fig. 9.

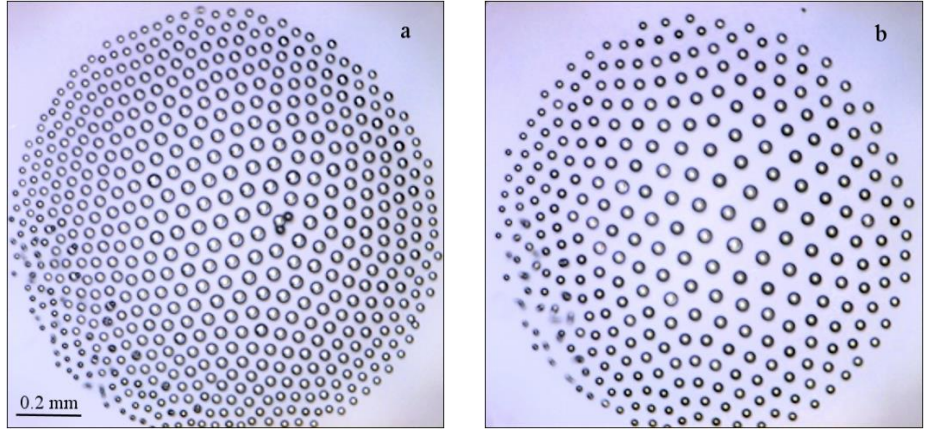

Fig. 9. Images of similar droplet clusters at different laser power close to that used in the present study: a – $W_L = 250\,\text{mW}$, b – $310\,\text{mW}$ (the scale bar is the same for both panels).

Possible explanation of a very small effect of electric field on distances between the particle is in the observed strong increase in the above mentioned strong effect of electric field on the growth rate of water droplets.

The regime of interaction between the steam flow and water droplets is mainly determined by the Reynolds number $\text{Re} = 2\rho u_0 R/\eta$, where $u_0$ is the maximum velocity of steam at the water layer surface and $\eta$ is the dynamic viscosity of steam. The calculations of [11] for maximum heat flux at the surface of water layer at $W_L = 280\,\text{mW}$ and $E = 0$ enable us to estimate the values of $u_0 = 0.08\,\text{m/s}$ and $\text{Re} = 2.1 \cdot 10^{-6}$. This very low Reynolds number is typical to the Stokes flow regime characterized by a linear momentum equation. Unfortunately, the spatial arrangement of water droplets in the cluster, and competitive effects of steam condensation and water evaporation, and diffusion in the steam-air mixture make the computational modeling of the gas flow and forces between the droplets very complicated. Such a modeling is at present not feasible. Therefore, our physical estimates are based on some relatively simple solutions.

The interaction between two identical spherical particles or droplets which are held fixed side by side against a uniform stream directed perpendicular to the line connecting the particles' centers was computationally analyzed in papers [30, 31]. The calculations for the range of $10 \leq \text{Re} \leq 150$ showed that the two spheres repel each other when the spacing is of the order of the diameter, and the repulsion is stronger at smaller spacing between the particles. On the other hand, the two particles weakly attract each other at the intermediate separation distances. The numerical results of [31] for $\text{Re} = 10$ and $50$ were slightly corrected in recent paper [32] and additional results for smaller distances between the particles showed a significant increase in the repelling force. The calculations of [31] at $\text{Re} = 5 \cdot 10^{-7}$ (in the Stokes regime) showed that the particles weakly repel each other at all separations. When the hydrodynamic interaction between the closely spaced particles becomes negligible, the so-called London–van der Waals attraction force should be considered [33, 34]. This physical limit is important for small particles suspended in a liquid, but it is not observed in the case of a droplet cluster.

There are several important differences between interaction of a uniform flow coming from a large distance to a couple of side by side particles and the flow around the droplet cluster. Perhaps, the most important difference is a small ratio of the cluster levitation height to the cluster diameter. A computational modeling the levitating cluster is definitely beyond the scope of the present paper. Therefore, our consideration is limited to a comparison of the Coulomb force between the neighboring charged water droplets and approximate value of the aerodynamic attraction force between the same droplets. The first of these forces is expressed as follows:

$$F_{el} = \frac{1}{4\pi\varepsilon_0} \frac{q_{dr}^2}{L^2} \qquad (5)$$

Equation (5) gives the value of $F_{el} = 8 \cdot 10^{-6}$ nN at $L = 80\,\mu\text{m}$ (we restrict ourselves by the pair Coulomb attraction of droplets). The aerodynamic force due to decrease in static pressure of a moving gas between the droplets can be estimated as:

$$F_{ad} = \frac{\rho(u_1^2 - u_0^2)}{2} \xi \pi R^2 \qquad (6)$$

where $u_1$ is the average velocity in the gap between the droplets, and $\xi < 1$ is a dimensionless coefficient. For the hexagonal structure of droplet cluster, the value of $u_1$ can be calculated as follows:

$$u_1 = \gamma u_0 \qquad \gamma = 1 \Big/ \left(1 - \frac{4\pi}{3\sqrt{3}} \frac{R^2}{L^2}\right) \qquad (7)$$

Bearing in mind that $R \ll L$, we obtain:

$$F_{ad} = \rho u_0^2 \frac{4\pi^2}{3\sqrt{3}} \xi \frac{R^4}{L^2} \qquad (8)$$

It is interesting that both $F_{el}$ and $F_{ad}$ are inversely proportional to $L^2$. At the same time, the absolute value of the aerodynamic force is in many orders of magnitude greater than the electrostatic force. Assuming $\rho = 0.65\,\text{kg/m}^3$, $u_0 = 0.08\,\text{m/s}$, $R = 20\,\mu\text{m}$, and $L = 80\,\mu\text{m}$ in (8), we obtain $F_{ad} = 20\,\mu\text{N}$ even at $\xi = 0.01$.

The above estimations confirm that aerodynamic forces are responsible for the interaction between the droplets in the levitating cluster. Of course, one needs much more sophisticated analysis to understand the observed spatial scale of the cluster regular pattern.

## 5. Conclusions

The analysis of laboratory experiments performed with an external electric field of various direction and intensity enabled the authors to obtain the following main results for a typical droplet cluster levitating over the locally heated water surface: (1) The external electric field leads to a significant increase in the rate of a condensational growth of water droplets; (2) The negative electric charge of single droplets in the central part of a cluster is equal to 473 elementary units of electric charge at the experimental conditions; (3) The polarization of water droplets results in an additional attraction force between the droplets and water layer. This force is comparable to the gravitational force and leads to early coalescence of small droplets with water layer; (4) The electrostatic force between the droplets is much less that the estimated aerodynamic force which is responsible for the cluster pattern. The results obtained are expected to be useful for further theoretical modeling of the phenomenon of levitating droplets clusters.

**Conflict of interests**

None declared.

**Acknowledgements**

The authors are grateful to the Russian Ministry of Education and Science (project no. 3.8191.2017/БЧ) for the financial support of the present study.

**References**

[1] A.A. Fedorets, Droplet cluster, JETP Lett. 79 (8) (2004) 372–374.

[2] A.A. Fedorets and L.A. Dombrovsky, Self-assembled stable clusters of droplets over the locally heated water surface: Milestones of the laboratory study and potential biochemical applications, Proc. 16$^{th}$ Int. Heat Transfer Conf. (IHTC-16), Aug. 10-15, 2018, Beijing, China, keynote paper IHTC16-KN17.

[3] T. Umeki, M. Ohata, H. Nakanishi, and M. Ichikawa, Dynamics of microdroplets over the surface of hot water, Sci. Rep. 5 (2015) paper 8046.

[4] A.A. Fedorets, M. Frenkel, E. Shulzinger, L.A. Dombrovsky, E. Bormashenko, and M. Nosonovsky, Self-assembled levitating clusters of water droplets: Pattern-formation and stability, Sci. Rep. 7 (2017) paper 1888.

[5] A.V. Shavlov, V.A. Dzhumandzhi, and S.N. Romanyuk, Electrical properties of water drops inside the dropwise cluster, Phys. Lett. A 376 (1) (2011) 39–45.

[6] A.V. Shavlov, V.A. Dzhumandzhi, and S.N. Romanyuk, Sound oscillation of dropwise cluster, Phys. Lett. A 376 (28-29) (2012) 2049–2052.

[7] A.A. Fedorets, Mechanism of stabilization of location of a droplet cluster above the liquid–gas interface, Tech. Phys. Lett. 38 (11) (2012) 988–990.

[8] L.A. Dombrovsky, A.A. Fedorets, and D.N. Medvedev, The use of infrared irradiation to stabilize levitating clusters of water droplets, Infrared Phys. Techn. 75 (2016) 124–132.

[9] A.A. Fedorets, M. Frenkel, E. Bormashenko, and M. Nosonovsky, Small levitating ordered droplet clusters: Stability, symmetry, and Voronoi entropy, J. Phys. Chem. Lett. 8 (22) (2017) 5599–5602.

[10] A.A. Fedorets, N.E. Aktaev, and L.A. Dombrovsky, Suppression of the condensational growth of droplets of a levitating cluster using the modulation of the laser heating power, Int. J. Heat Mass Transfer 127 (part A) (2018) 660–664.

[11] A.A. Fedorets and L.A. Dombrovsky, Generation of levitating droplet clusters above the locally heated water surface: A thermal analysis of modified installation, Int. J. Heat Mass Transfer 104 (2017) 1268–1274.

[12] Z. Chen, The Finite Element Method: Its Fundamentals and Applications in Engineering, World Sci. Publ., Singapore, 2011.

[13] O.C. Zienkiewicz, R.L. Taylor, and J.Z. Zhu, The Finite Element Method: Its Basis and Fundamentals, Seventh edition, Elsevier, New York, 2013.

[14] N. Jonassen, Electrostatics, Second edition, Kluver, Norwell, MA, 2002.

[15] A.A. Fedorets, On the mechanism of non-coalescence in a drop cluster, JETP Lett. 81 (9) (2005) 437–441.

[16] E.A. Arinstein and A.A. Fedorets, Mechanism of energy dissipation in a droplet cluster, JETP Lett. 92 (10) (2010) 658–661.

[17] A.A. Fedorets, I.V. Marchuk, and O.A. Kabov, On the role of capillary waves in the mechanism of coalescence of a droplet cluster, JETP Lett. 99 (5) (2014) 266–269.

[18] W.A. Sirignano, Fluid Dynamics and Transport of Droplets and Sprays, Cambridge: Cambridge Univ. Press, 1999.

[19] L.A. Dombrovsky and S.S. Sazhin, A simplified non-isothermal model for droplet heating and evaporation. Int. Comm. Heat Mass Transfer 30 (6) (2003) 787–796.

[20] A.A. Fedorets, L.A. Dombrovsky, and P.I. Ryumin, Expanding the temperature range for generation of droplet clusters over the locally heated water surface, Int. J. Heat Mass Transfer 113 (2017) 1054–1058.

[21] Y. Asakawa, Promotion and retardation of heat transfer by electric fields, Nature, 261 (1976) 220–221.

[22] A. Wolny and R. Kaniuk, The effect of electric field on heat and mass transfer, Drying Techn. 14 (2) (1996) 195–216.

[23] F.C. Lai and K.-W. Lai, EHD-enhanced drying with wire electrode, Drying Techn. 20 (7) (2002) 1393–1405.

[24] Y. Okuno, M. Minagawa, H. Matsumoto, and A. Tanioka, Simulation study on the influence of an electric field on water evaporation, J. Molecular Structure: THEOCHEM 904 (1-3) (2009) 83–90.

[25] B. Kamkari and A.A. Alemrajabi, Investigation of electrodynamically-enhanced convective heat and mass transfer from water surface, Heat Transfer Eng. 31 (2) (2010) 138–146.

[26] R. Reiter, Fields, Currents and Aerosols in the Lower Troposphere, CRC Press, New York, 1986.

[27] Y. Dong and J. Hallett, Charge separation by ice and water drops during growth and evaporation, J. Geophys. Res. D. Atmos. 97 (D18) (1992) 20361–20371.

[28] R.A. Black and J. Hallett, Electrification of the hurricane, J. Atmos. Sci. 56 (12) (1999) 2004–2028.

[29] R.G. Harrison and K.S. Carslaw, Ion-aerosol-cloud processes in the lower atmosphere, Rev. Geophys. 41 (3) (2003) paper 1012.

[30] I. Kim, S. Elgobashi, and W.A. Sirignano, Three-dimensional flow over two spheres placed side by side, J. Fluid Mech. 246 (1993) 465–488.

[31] R. Folkersma, H.M. Stein, and F.N. van de Vosse, Hydrodynamic interaction between two identical spheres held fixed side by side against a uniform stream directed perpendicular to the line connecting the spheres' centres, Int. J. Multiphase Flow 26 (5) (2000) 877–887.

[32] T. Kotsev, Viscous flow around spherical particles in different arrangements, MATEC Web of Conf. 145 (2018) paper 03008.

[33] H.C. Hamaker, The London–van der Waals forces between spherical particles, Physica 4 (10) (1937) 1058–1072.

[34] J.N. Israelashvili, Van der Waals forces in biological systems, Quart. Rev. Biophys. 6 (4) (1973) 341–387.